\documentstyle[amssymb,aps,epsfig,twocolumn]{revtex}

\begin{document}

\title{Robust quantum gates on neutral atoms with cavity-assisted photon-scattering}
\author{L.-M. Duan$^{1}$, B. Wang$^{1}$, H. J. Kimble$^{2}$}
\address{$^{1}$ FOCUS Center and MCTP, Department of Physics,
University of Michigan, Ann Arbor, MI 48109-1120\\
$^{2}$Norman Bridge Laboratory of Physics 12-33, California
Institute of Technology, Pasadena, CA 91125} \maketitle
\begin{abstract}
We propose a scheme to achieve quantum computation with neutral
atoms whose interactions are catalyzed by single photons.
Conditional quantum gates, including an $N$-atom Toffoli gate and
nonlocal gates on remote atoms, are obtained through
cavity-assisted photon scattering in a manner that is robust to
random variation in the atom-photon coupling rate and which does
not require localization in the Lamb-Dicke regime. The dominant
noise in our scheme is automatically detected for each gate
operation, leading to signalled errors which do not preclude
efficient quantum computation even if the error probability is
close to the unity.

\textbf{PACS numbers: 03.67.Lx, 03.67.Hk, 42.50.-p}
\end{abstract}

Neutral atoms in optical cavities have been one of the pioneering avenues
for the implementation of quantum computation and networking \cite
{1,turchette95,2,3}. Nevertheless, the experimental requirements associated
with these approaches turn out to be very challenging. In particular,
although significant experimental advances have been reported recently in
transmitting and trapping single atoms in high finesse cavities \cite
{3,mckeever03,8,maunz04,mckeever04,boca04,maunz05,keller04,kreuter04}, no
experiment has yet achieved a well defined number of atoms $N\geq 2$ each of
which is strongly coupled to the cavity mode, individually addressable, and
localized to the Lamb-Dicke limit, as is required for the protocol of Ref.
\cite{1}. To realize a more scalable system, Chapman \textit{et al.}
proposed an architecture in which a transverse optical lattice is employed
to translate atoms into and out of a high-finesse cavity for entangling gate
operations \cite{8}. Transport that preserves internal state coherence has
been demonstrated for both ions \cite{rowe02} and atoms \cite{kuhr03}.
However, although the approach of Ref. \cite{8} does solve the problem of
separate addressing of many atoms in a tiny cavity, there remain significant
obstacles to achieving Lamb-Dicke confinement \cite{example} and strong
coupling for any scheme that has yet been proposed.

To overcome these difficulties and to provide several new
capabilities for quantum logic, in this paper we propose a scheme
for atomic quantum gates whereby atom-atom interactions are
catalyzed by single photons in a fashion that is robust to various
sources of practical noise. More specifically, a controlled
phase-flip gate between two atoms is achieved by cavity-assisted
scattering of a single-photon pulse from the cavity in which the
atoms are localized \cite{9}. This gate is insensitive to
uncertainties in the atom-photon coupling rate, thereby obviating
the requirement for Lamb-Dicke localization. It is also robust to
all sources of photon loss, including, for instance, atomic
spontaneous emission, photon collection and detection
inefficiency, and any vacuum component in the scattering pulse.
Such noise is automatically detected for each gate, leading to a
finite failure probability of the gate operation. As shown in
Refs. \cite{10,11}, efficient quantum computation can nevertheless
be achieved even if the associated failure probability is close to
unity. Moreover, our scheme can be readily extended to achieve a
Toffoli gate for $N$ atoms in a single step and to realize
nonlocal gates on remote atoms trapped in different cavities. The
direct $N$-bit gate could lead to more efficient construction of
quantum circuits, and the nonlocal gates on remote atoms naturally
integrates local computation with quantum networking.

To explain the idea of the gate operation, we first consider two atoms in a
single-sided cavity. To have a scalable architecture, one can follow Ref.
\cite{8} to assume there are transverse optical lattice potentials to move
the target atoms into and outside the cavity \cite{note,schrader04}. Each
atom has three relevant levels as shown in Fig. 1. The qubit is represented
by different hyperfine levels $\left\vert 0\right\rangle $ and $\left\vert
1\right\rangle $ in the ground-state manifold. The atomic transition from $%
\left\vert 1\right\rangle $ to an excited level $\left\vert e\right\rangle $
is resonantly coupled to a cavity mode $a_{c}$. The state $\left\vert
0\right\rangle $ is decoupled due to the large hyperfine splitting.

\begin{figure}[tb]
\epsfig{file=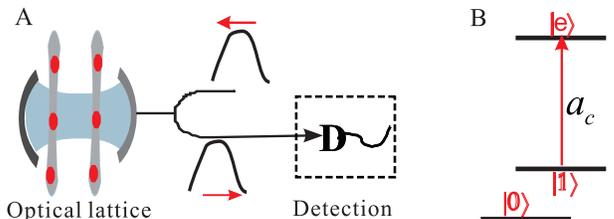,width=8cm} %\includegraphics[width=8.6cm]{d70p1.eps}
\caption{(A) Schematic setup for implementation of the controlled phase flip
(CPF) gate on two atoms inside the cavity through the photon-scattering
interaction. Any pair of atoms can be transmitted into the cavity for a
collective gate operation through a transverse optical lattice potential as
suggested in Ref. \protect\cite{8} \protect\cite{12}. For a more robust
implementation of the gate, we add a single-photon detector to detect the
output photon pulse as illustrated inside the dashed box. (B) The relevant
level structure of the atoms and the coupling configuration.}
\label{fig1}
\end{figure}

To perform a collective quantum gate on the two atoms, we reflect a
single-photon pulse from the cavity. This single-photon pulse, with its
state denoted as $\left| p\right\rangle $, is resonant with the bare cavity
mode $a_{c}$. If the photon pulse is sufficiently long (with its bandwidth $%
\Delta \Omega $ much smaller than the cavity decay rate $\kappa $),
reflection of the pulse from a resonant cavity absent an atom will leave the
pulse shape almost unchanged but will flip its global phase, as we later
characterize in detail. For the case that both of the atoms are in the $%
\left| 0\right\rangle $ state, this is precisely the nature of the resonant
reflection since there is negligible atom-cavity coupling and hence no shift
of the resonant frequency of the cavity mode. After reflection, the
atom-photon state $\left| 0\right\rangle _{1}\left| 0\right\rangle
_{2}\left| p\right\rangle $ evolves into $-\left| 0\right\rangle _{1}\left|
0\right\rangle _{2}\left| p\right\rangle $, where the subscripts $1,2$
denote the two intracavity atoms. However, if either or both of the atoms
are in the state $\left| 1\right\rangle $, the effective frequency of the
dressed cavity mode will be shifted due to the atom-cavity coupling, which
is described by the Hamiltonian
\begin{equation}
H=\hbar \sum_{i=1,2}g_{i}\left( |e\rangle _{i}\langle 1|a_{c}+|1\rangle
_{i}\langle e|a_{c}^{\dag }\right) .  \label{H}
\end{equation}
If the coupling rates $g_{i}\gg (\Delta \Omega ,\kappa ,\gamma _{s})$, where
$\gamma _{s}$\ is the rate of spontaneous decay of $|e\rangle $, then the
frequency shift will have a magnitude comparable with $g_{i}$, so that the
incident single-photon pulse will be reflected by an off-resonant cavity.
Hence, both the shape and global phase will remain unchanged for the
reflected pulse. Due to this property, the component states $\left|
0\right\rangle _{1}\left| 1\right\rangle _{2}\left| p\right\rangle ,$ $%
\left| 1\right\rangle _{1}\left| 0\right\rangle _{2}\left| p\right\rangle ,$
and $\left| 1\right\rangle _{1}\left| 1\right\rangle _{2}\left|
p\right\rangle $ are likewise unaffected by reflection process. The net
effect of these two subprocesses is that the reflection of a single-photon
pulse from the cavity actually performs a controlled phase-flip gate (CPF) $%
U_{12}=\exp \left( i\pi \left| 00\right\rangle _{12}\left\langle 00\right|
\right) $ on the two atoms while leaving the photon state unchanged
(unentangled). Hence, in the ideal case the reflected photon can be utilized
to catalyze subsequent gate operations.

However, in a realistic setting our scheme can be performed in a more robust
fashion by detecting the output pulse with a single-photon detector. By this
means, gate errors due to all sources of photon loss, including atomic
spontaneous emission, cavity mirror absorption and scattering, imperfection
in the photon source, and photon collection and detection inefficiencies,
are always signaled by the absence of a photon count. As a result, these
dominant sources of noise only lead to probabilistic signaled errors, which
yield a finite failure probability of the gate but which have no
contribution to the gate infidelity if the operation succeeds (i.e., if a
photon count is registered). For this class of errors, efficient quantum
computation is possible with an arbitrarily small gate success probability $%
p $ \cite{10}. Compared with deterministic gates, the required extra
computational overhead due to the small gate success probability $p$ scales
efficiently (polynomially) both with $1/p$ and the computational scale
characterized by the number of qubits $n$ \cite{10}. Because of this
robustness, the input single-photon pulse can also be replaced by a simple
weak coherent pulse $\left\vert \alpha \right\rangle $ with the mean photon
number $\left\vert \alpha \right\vert ^{2}\ll 1$. This replacement does not
give any essential problem in terms of scaling, although the individual gate
efficiency (the success probability) is indeed significantly reduced by a
factor of $\left\vert \alpha \right\vert ^{2}$.

Before going to the detailed theoretical characterization of the gate
fidelity and efficiency, we next present some extensions of the above
scheme. First, our scheme can be readily extended to perform a Toffoli gate
on $N$ atoms in a single time step. If one reflects a single-photon pulse
from a\ cavity with $N$ atoms trapped inside, the pulse will have a flip of
its global phase if and only if all the atoms are in the $\left\vert
0\right\rangle $ state. So, this reflection performs a Toffoli gate $%
U_{12\cdots N}=\exp \left( i\pi \left\vert 00\cdots 0\right\rangle
_{12\cdots N}\left\langle 00\cdots 0\right\vert \right) $ on all the atoms
while leaving the photon state unentangled. This direct $N$-bit gate could
lead to more efficient construction of circuits for quantum computation. For
instance, the reflection operation in the Grover's search algorithm can be
realized in a single step with the $N$-bit Toffoli gate \cite{12}.

Second, the above scheme can also be extended to perform nonlocal
gates on two remote atoms trapped in different cavities, as
illustrated in Fig. 2. For this purpose, one uses a single-photon
(or weak-coherent) pulse which is in an equal superposition state
$\left( \left| H\right\rangle +\left| V\right\rangle \right)
/\sqrt{2}$ of the $H$ and $V$ polarization components. With a
polarization beam splitter (PBS1), the $H$ and $V$ components of
the pulse are ``bounced''\ back from the atom-cavity system and a
mirror \textit{M}, respectively, with the reflection from
\textit{M} leaving the incident pulse unchanged. The overall
reflection from the cavity and the mirror \textit{M} actually
performs the gate operation $U_{1p}=\exp \left( i\pi \left|
0H\right\rangle _{1p}\left\langle 0H\right| \right) $ on atom $1$
and the photon pulse $p$, so that there is a phase flip only when
the atom is in the state $\left| 0\right\rangle $ and the photon
is in the polarization $\left| H\right\rangle $ \cite{9}. The
pulse is reflected successively from the two cavity setups, with a
quarter-wave plate (QWP1) inserted into the optical path between
the two reflections which performs a Hardmard rotation on the
photon's polarization $\left| H\right\rangle
\rightarrow \left( \left| H\right\rangle +\left| V\right\rangle \right) /%
\sqrt{2},$ $\left| V\right\rangle \rightarrow \left( \left| V\right\rangle
-\left| H\right\rangle \right) /\sqrt{2}$. The photon is detected by two
single-photon detectors D1 and D2 after the reflections, corresponding to a
measurement of its polarization in the basis $\left( \left| V\right\rangle
\pm \left| H\right\rangle \right) /\sqrt{2}$ (after the QWP2 and the PBS3;
see Fig. 2). For a detection event in D2, a phase flip operation $\sigma
_{1}^{z}$ is performed on the atom 1, while no operation is applied if D1
clicks. The net effect of these operations is the desired CPF gate $%
U_{12}=\exp \left( i\pi \left| 00\right\rangle _{12}\left\langle 00\right|
\right) $ on the two remote atoms $1,2$. Among other applications, this
nonlocal gate and its extension to multiple atom-cavity systems provide a
convenient avenue for quantum networking. As before for the case of a single
cavity, in this distributed setting any noise leading to photon loss is
always signaled by the absence of a photon count from either D1 or D2.

\begin{figure}[tb]
\epsfig{file=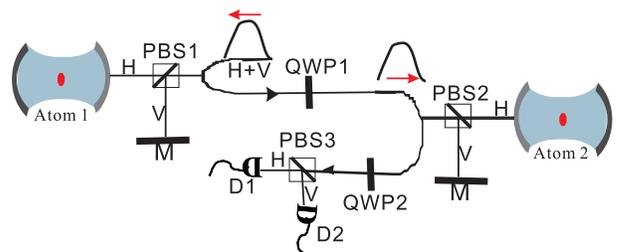,width=8cm} %\includegraphics[width=8.6cm]{d70p1.eps}
\caption{Schematic of the setup for implementation of nonlocal gates on two
atoms 1 and 2 trapped in distant cavities. Not shown are circulators (e.g.,
Faraday devices) to redirect the output beams along paths distinct from the
inputs. See the text for further explanation.}
\label{fig2}
\end{figure}

We now present a more detailed theoretical model of our scheme and
characterize the influence of some practical sources of noise. The input
single-photon pulse with a normalized shape function $f_{in}\left( t\right) $
and a duration $T$ can be described by the state $\left| p\right\rangle
=\int_{0}^{T}f_{in}\left( t\right) a_{in}^{\dagger }\left( t\right) dt\left|
\text{vac}\right\rangle $, where $\left| \text{vac}\right\rangle $ denotes
the vacuum state and $a_{in}^{\dagger }\left( t\right) $ is the
one-dimensional optical field operator with the commutation relation $\left[
a_{in}\left( t\right) ,a_{in}^{\dagger }\left( t^{\prime }\right) \right]
=\delta \left( t-t^{\prime }\right) $ \cite{13}. The cavity mode $a_{c}$ is
driven by the input field $a_{in}\left( t\right) $ through the Langevin
equation \cite{13}
\begin{equation}
\dot{a}_{c}=-i[a_{c},H]-\left( \kappa /2\right) a_{c}-\sqrt{\kappa }%
a_{in}\left( t\right) \text{ ,}
\end{equation}
where $\kappa $ is the cavity decay rate and the Hamiltonian $H$ is given in
Eq. (1) for the case of two atoms; generalization to multiple atoms is
straightforward. To account for atomic spontaneous emission with a rate $%
\gamma _{s}$, we add an effective term $\left( -i\gamma _{s}\right) \left|
e\right\rangle \left\langle e\right| $ to the Hamiltonian $H$. The output
field $a_{out}\left( t\right) $ of the cavity is connected with the input
through the input-output relation $a_{out}\left( t\right) =a_{in}\left(
t\right) +\sqrt{\kappa }a_{c}$.

The final atom-photon state can be numerically solved from the above set of
equations through discretization of the continuum optical fields (for
details on the numerical method, see Refs. \cite{9,14}). We use the
following two quantities to characterize the imperfections in our scheme.
(1) Due to various sources of photon loss, photons in the cavity may be lost
with then no photon count at the detectors. Hence, we calculate the success
probability of a photon count at the detector to characterize the efficiency
of the scheme. (2) Even if a photon emerges, there may still be
imperfections of the atomic gate mainly due to the shape distortion of the
photon pulse after reflection from the cavity, which can be characterized
through the gate fidelity. Without loss of the photon, the final atom-photon
state can be written as $\left| \Psi _{out}\right\rangle
=\sum_{i_{1}i_{2}}c_{i_{1}i_{2}}\left| i_{1}i_{2}\right\rangle _{a}\left|
p\right\rangle _{i_{1}i_{2}}$, where $\sum_{i_{1}i_{2}}c_{i_{1}i_{2}}\left|
i_{1}i_{2}\right\rangle _{a}$ ($i_{1},i_{2}=0,1$) is the general form for
the input state of the two atoms. The output photon state $\left|
p\right\rangle _{i_{1}i_{2}}$ corresponds to the atomic component $\left|
i_{1}i_{2}\right\rangle _{a}$, and is given by $\left| p\right\rangle
_{i_{1}i_{2}}=\int_{0}^{T}f_{i_{1}i_{2}}^{out}\left( t\right)
a_{out}^{\dagger }\left( t\right) dt\left| \text{vac}\right\rangle $ with a
shape $f_{i_{1}i_{2}}^{out}\left( t\right) $. Ideally, the output state $%
\left| \Psi _{out}^{id}\right\rangle $ would have the shape functions $%
f_{00}^{out}\left( t\right) =-f_{in}\left( t\right) $ and $%
f_{i_{1}i_{2}}^{out}\left( t\right) =f_{in}\left( t\right) $ (for $%
i_{1},i_{2}\neq 0$), which realizes a perfect CPF gate $U_{12}$ on the
atoms. Hence to characterize the gate imperfection, we calculate the
fidelity $F\equiv \left| \left\langle \Psi _{out}^{id}\right. \left| \Psi
_{out}\right\rangle \right| ^{2}$, which is directly extendable to any
number of atoms. In the following calculation of the fidelity $F$, we choose
the input state $\left[ \left( \left| 0\right\rangle +\left| 1\right\rangle
\right) /\sqrt{2}\right] ^{\otimes N}$ for the case of $N$ atoms.

The results from our calculations are summarized in Fig. 3. First, Fig. 3A
shows the component pulse shape $f_{i_{1}i_{2}}^{out}\left( t\right) $
corresponding to a Gaussian input $f_{in}\left( t\right) $ for the case of
two atoms. Only the component $f_{00}^{out}\left( t\right) $ has a notable
phase distortion; all others are basically indistinguishable from the input.
To account for random variation in the coupling rates $g_{i}$, we have also
calculated $f_{i_{1}i_{2}}^{out}\left( t\right) $ for $g_{i}$ varying from $%
2\kappa $ to $6\kappa $. The output pulse shapes are nearly identical for $%
g_{i}$ varying in this range, which is typical of current experiments \cite
{3,mckeever03,8,maunz04,mckeever04,boca04,maunz05}. Figure 3B shows the
corresponding fidelity $F$ of the CPF (or Toffoli) gate from the shape
distortion noise with the atom number $N=2,3,4,5$. The fidelity $F$ improves
with increase of the pulse duration $T$ since the shape distortion is
reduced for longer pulses. $F$ also increases with the atom number $N$,
which is a bit surprising but actually reasonable: for the $N$-atom state $%
\left[ \left( \left\vert 0\right\rangle +\left\vert 1\right\rangle \right) /%
\sqrt{2}\right] ^{\otimes N}$, the fraction of the component
$\left\vert 0\right\rangle ^{\otimes N}$ goes down as $1/2^{N}$,
and the pulse shape distortion noise comes dominantly from this
component. Because the component $\left\vert 0\right\rangle
^{\otimes N}$ dominates the contribution to the gate infidelity,
$F$ is also very insensitive to variation of the coupling rates
$g_{i}$. We have verified that there is no notable change of $F$
($\delta F<10^{-4}$) in Fig. 3B for $g_{i}$ varying from $2\kappa
$ to $6\kappa $.

Any source of photon loss has no contribution to the gate fidelity
but instead influences gate efficiency (success probability). A
fundamental source of photon loss is atomic spontaneous emission.
Figure 3C shows the failure probability $P_{\text{sp}}$ of the
gate due to this source of noise,
with the noise rate $\gamma _{s}=\kappa $. For $N$ atoms with equal $g_{i}=g$%
, the probability $P_{\text{sp}}$ can be well fit by an empirical formula $%
P_{\text{sp}}\approx P_{\text{emp}}\equiv \sum_{n=1}^{N}\left(
N!/n!(N-n)!2^{N}\right) \left[ 1+ng^{2}/\kappa \gamma _{s}\right]
^{-1}$. The empirical $P_{\text{emp}}$ can be understood as a
probability averaged over all the Dicke-state components in the
input state $\left[ \left( \left\vert 0\right\rangle +\left\vert
1\right\rangle \right) /\sqrt{2}\right] ^{\otimes N},$ with the
$n$th Dicke-component having an effective coupling rate
$\sqrt{n}g$ to the cavity mode. We have also simulated the loss
probability $P_{\text{sp}}$ when the coupling rates $g_{i}$ are
different and vary during the gate operation, for instance, as
would be caused by the atoms' thermal motion. With some typical
choice of the relevant experimental parameters, the result is
shown in Figure 3D, which is qualitatively similar
to the constant coupling rate case with an effective average over $%
\left\vert g_{i}\right\vert $. Other sources of photon loss can be similarly
characterized. For instance, with a finite photon collection and detection
efficiency $\eta $, the success probability of each gate will be simply
reduced by a factor of $\eta $.

In summary, we have proposed a new scheme for robust atomic gates by way of
interactions mediated by cavity-assisted photon scattering. These gates are
robust to all sources of photon loss that are typically the dominant source
of noise in experimental implementations, and are furthermore insensitive to
randomness in the coupling rates caused by fluctuations in atomic position.
Beyond two-atom gates illustrated in Fig. 1, our scheme can also be employed
for realization of an $N$-atom Toffoli gate in a single step and for the
implementation of nonlocal gates on distant atoms as in Fig. 2. We have
characterized the efficacy of our scheme through exact numerical simulations
that incorporate various sources of experimental noise. These results
demonstrate the practicality of our scheme by way of current experimental
technology.

\begin{figure}[tb]
\epsfig{file=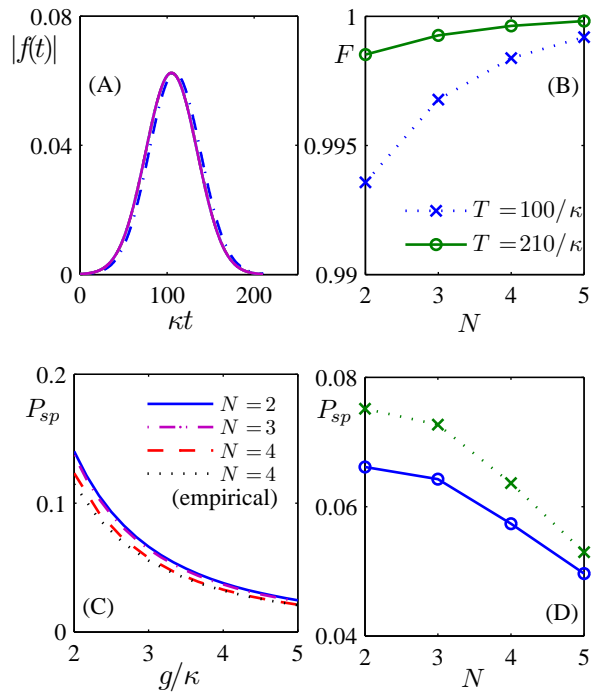,width=8cm}
\caption{(A) The shape functions $\left\vert f\left( t\right)
\right\vert $ for the input pulse (solid curve) and the reflected
pulses with the atoms in different component states $\left\vert
i_{1}i_{2}\right\rangle _{a}$. The shape function for the atom in
the state $\left\vert 00\right\rangle _{a}$ is shown by the
dash-dot curve. With the coupling rate $g$ in a typical range from
$2\protect\kappa $ to $6\protect\kappa $, the shape functions for
the atoms in all the other component states are indistinguishable
from that of the input pulse (the solid curve). We have assumed a
Gaussian shape for the input pulse with $f_{in}\left( t\right)
\propto \exp \left[ -\left( t-T/2\right) ^{2}/\left( T/5\right)
^{2}\right] $ , where $t$ ranges from $0$ to $T$ and
$T=210/\protect\kappa $ for this example. (B) The gate fidelity
versus the number of atoms with the pulse duration
$T=100/\protect\kappa $ (the dotted curve)
and $T=210/\protect\kappa $ (the solid curve), respectively. (C) The photon loss probability $%
P_{sp}$ due to atomic spontaneous emission shown as a function of
the coupling rate $g$ in units of $\protect\kappa $ with the atom
number $N=2,3,4$. The dotted curves shows $P_{sp}$ calculated from
the empirical formula given in the text for $N=4$. (D) Comparison
of the photon loss $P_{sp}$ for a constant coupling rate
$g=3\protect\kappa $ (the solid curve)) and for a time varying
rate $g_{i}(t)=3\protect\kappa (1+\sin (\protect\nu t+\protect\phi
_{i})/3)$ (the dotted curve) for the ith atom, where $\protect\nu
=\protect\kappa/6 $ corresponds to a
typical atom's axial oscillation frequency in the trap, and $\protect\phi %
_{i}$ are taken as random numbers accounting for the atoms' random
initial positions. $g_{i}(t)$ is chosen so that its maximum and
minimum differ by a factor of $2$, which exceeds that in current
experiments \protect\cite
{boca04}. Other parameters for Figs. (A) and (B) are $\protect\gamma _{s}=%
\protect\kappa $ and $g=3\protect\kappa $, and for Figs. (C) and (D), $%
\protect\gamma _{s}=\protect\kappa $ and $T=210\protect\kappa $.}
\label{fig3}
\end{figure}

We gratefully acknowledge the contributions of W. D. Phillips,
whose question stimulated these investigations. We thank also S.
Lloyd for helpful discussions. This work was supported by the
National Science Foundation (Awards Nos. 0431476, EIA-0086038, and
PHY-0140355), the Advanced Research and Development Activity under
ARO contracts, the A. P. Sloan Fellowship (LMD), and the Caltech
MURI Center for Quantum Networks (HJK).


\begin{thebibliography}{99}
\bibitem{example}  For example, if a far-off resonance dipole-force trap
(FORT) is utilized for conveying an atom along a direction transverse to the
cavity axis \cite{8,kuhr03} and if an atom is cooled to the motional ground
state of the FORT, the spread of the atomic wave packet along the direction
of the cavity axis is typically still comparable to the optical wavelength
for the cavity QED field due to focusing the FORT beams within the
constraints of the cavity geometry. This causes significant uncertainty in
the atom-photon coupling rate.
\bibitem{note}  The scheme also applies to the more traditional cavity
quantum computation architecture with an array of atoms in the cavity.
Separate addressing of different atoms could be achieved through frequency
selection either with Stark shifts from focused laser beams or with
gradients from effective magnetic fields \cite{schrader04}.
\end{thebibliography}
\end{document}